\documentclass[osajnl,twocolumn,showpacs,superscriptaddress,10pt]{revtex4-1}
\usepackage{amsmath,amssymb,graphicx,epstopdf}

\newcommand{\be}{\begin{equation}}
\newcommand{\ee}{\end{equation}}
\newcommand{\bea}{\begin{eqnarray}}
\newcommand{\eea}{\end{eqnarray}}

\newcommand{\la}{\langle}
\newcommand{\ra}{\rangle}
\newcommand{\lb}{\left[}
\newcommand{\rb}{\right]}
\newcommand{\lp}{\left(}
\newcommand{\rp}{\right)}
\newcommand{\bolds}[1]{{\boldsymbol #1}}

\def\nn{\nonumber\\}

\begin{document}

\title{Single-exposure profilometry using partitioned aperture wavefront
imaging}

\author{Roman Barankov} \email{Corresponding author: barankov@bu.edu}
\author{Jerome Mertz}
\affiliation{Department of Biomedical Engineering, Boston University, 44
Cummington Street, Boston, Massachusetts 02215, USA}

\begin{abstract}
We demonstrate a technique for instantaneous measurements of surface
topography based on the combination of a partitioned aperture wavefront imager
with a standard lamp-based reflection microscope. The technique can operate at
video rate over large fields of view, and provides nanometer axial resolution
and sub-micron lateral resolution. We discuss performance characteristics of
this technique, which we experimentally compare with scanning white light
interferometry.
\end{abstract}


\maketitle

Optical surface profiling is important in many applications ranging from
precision optics to semiconductor industry. Currently, it is dominated by
scanning white light interferometry (SWLI). In this approach, a surface
profile is recovered from interferometric intensity variations recorded at a
camera as a function of relative pathlength differences between sample and
reference
beams~\cite{Flournoy_1972,Davidson_1987,Kino_1990,Wyant_1992,Wyant_2013}. The
lateral resolution of the method is defined by the diffraction limit, which
for visible light is a fraction of a micron. The axial resolution, approaching
sub-nanometer range, is limited by noise in the system, such as shot noise,
detector noise, or mechanical uncertainties in the relative pathlengths of the
beams. In particular, the requirement of scanning in SWLI exacerbates the
problem of mechanical uncertainties and limits applications to quasi-static
profiling. Practical implementations of SWLI require precision mechanics which
add to the cost of commercial devices.

In this work we describe a method of surface profiling that provides
instantaneous measurements with resolution and dynamic range comparable to
those of SLWI, yet characterized by a simple, robust and inexpensive design.
The key element of our method is a Partitioned Aperture Wavefront (PAW)
imager~\cite{PAW}, which is a passive add-on that can be incorporated into any
standard widefield microscope. PAW imaging provides simultaneous phase and
amplitude contrast that is quantitative. It has the advantages of being fast
(single exposure), achromatic (works with lamp or LED illumination), and light
efficient (works with extended sources). We previously demonstrated an
implementation of PAW imaging in a transmission microscope configuration to
measure the phase shifts induced by biological cells on a slide~\cite{PAW}.
Here, we implement PAW imaging in a reflection microscopy configuration to
measure surface topography. We describe the performance of this device, which
we compare with SWLI.

\begin{figure}[htbp]
\centerline{\includegraphics[width=1.\columnwidth]{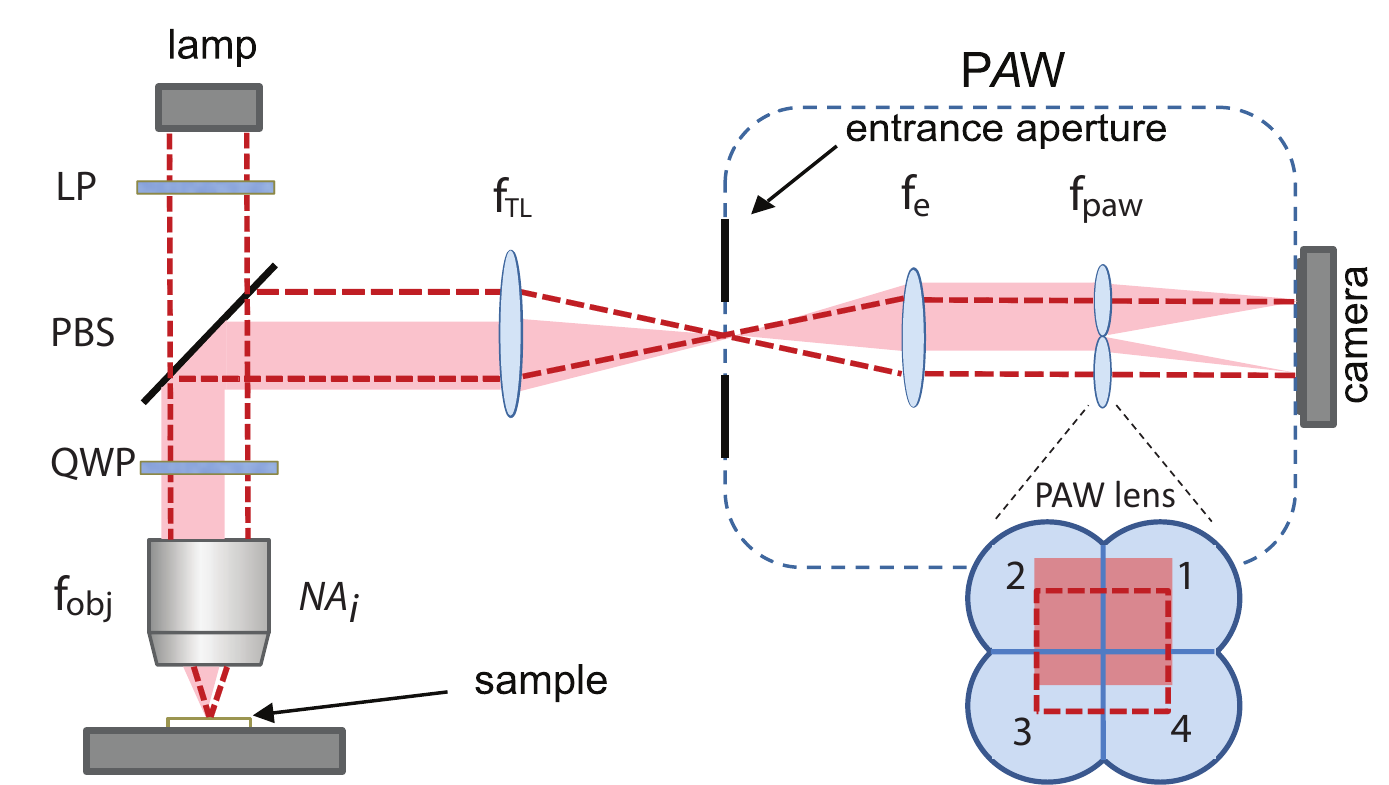}}
\caption{(color online) Experimental setup: $f_{\rm{obj}}$, $f_{\rm{TL}}$,
$f_e$, and $f_{\rm{paw}}$ denote a microscope objective and lenses. A linear
polarizer (LP), quarter-wave plate (QWP), and polarizing beam splitter (PBS)
minimize spurious back-reflections. The 3f PAW module includes a composite
lens (inset) that projects four oblique-detection images onto a camera.}
\label{fig:setup}
\end{figure}

PAW imaging is based on partitioning the detection aperture of a standard
imaging device into four quadrants, similarly to pyramidal wavefront
imaging~\cite{Iglesias_2011} but with the difference that the partitioning is
performed by four off-axis achromatic lenses rather than four prisms. These
lenses provide four oblique-detection images that are simultaneously acquired
with a single camera. The different perspectives presented by the four images
enable the reconstruction of wavefront phase and amplitude with a simple
numerical algorithm that runs in real time (here video rate).

Figure~\ref{fig:setup} illustrates an implementation of PAW imaging in a
reflection configuration suitable for surface profilometry. This
implementation consists of a standard reflection microscope based on K\"ohler
illumination with an illumination numerical aperture $\rm{NA}_i$. The
microscope camera, normally located at the intermediate image plane, is set
back to allow the insertion of the PAW module. This relays the intermediate
image through a 3f imaging system comprising an entrance lens of focal length
$f_e$ and the composite PAW lens. The latter consists of four off-axis lenses
of focal length $f_{\rm{paw}}$, cut and glued together in a quatrefoil
geometry (see Fig.~\ref{fig:setup} inset). A square entrance aperture is
placed in the intermediate image plane to prevent overlapping of the four
images at the camera plane. This square aperture defines the system field of
view (FOV).

The principle of surface profilometry based on PAW imaging can be understood
from simple geometrical optics (see Fig.~\ref{fig:setup}). Let us first
consider a flat reflective sample orthogonal to the optical axis. Provided the
illumination source is symmetrically distributed about the optical axis, the
illumination rays incident on the sample are, on average, normal to the sample
plane. The reflected rays are thus also, on average, normal to the sample
surface, and the power through the four PAW lenses is distributed equally
(dashed lines). That is, the four recorded images are of equal uniform
intensity. Let us now consider a local slope in the sample profile. This leads
to a local off-axis tilting of the reflected rays, and hence to an unbalancing
of the power distribution through the PAW lenses (shaded). The local intensity
differences in the recorded images thus encode local slope variations in the
sample surface profile.

These considerations, applicable in the paraxial approximation, suggest the
basic reconstruction algorithm for surface profiling. Specifically, in the
case of a square illumination aperture used in our work, the local tilt angles
along the transverse axes are defined by the simple algebraic combinations
~\cite{Iglesias_2011,PAW}:
\bea\label{eq:tilts}
\theta_x&=&{\rm{NA}_{i}}\lp
I_1-I_2-I_3+I_4\rp/I_{\rm tot},\nn
\theta_y&=&{\rm{NA}_{i}}\lp
I_1+I_2-I_3-I_4\rp/I_{\rm tot},
\eea
where $I_k$ are the image intensities recorded in the four quadrants of the
camera (see Fig.~\ref{fig:setup}), and $I_{\rm tot}=\sum_{k=1}^4 I_k$ is the
total intensity, equivalent to a standard widefield image of the sample.
Provided the detection numerical aperture $\rm{NA}_d$ is at least twice
$\rm{NA}_i$, then Eqs.~(\ref{eq:tilts}) are accurate for tilt angles
$|\theta_{x,y}|\le \rm{NA}_i$, characterizing the dynamic range of the tilt
measurements.

Physically, the tilt angles~(\ref{eq:tilts}) encode surface gradients,
\bea\label{eq:tilt_gradients}
\theta_x=2\nabla_x h,\quad\theta_y=2\nabla_y h,
\eea
where $h=h(\bolds{\rho})$ is the surface profile at the position
$\bolds{\rho}=(x,y)$ in the sample, and the factor of two accounts for the
doubling of the reflected angle with respect to the surface normal.

Surface profile is reconstructed by integration of
Eq.~(\ref{eq:tilt_gradients}), which we perform using a spiral phase Fourier
integration method described in Ref.~\cite{Arnison_2004}. In the continuum
limit we have
\bea\label{eq:h_basic}
h(\bolds{\rho} )-h_0 &=&\frac{1}{4\pi i}\int d^2\bolds{\kappa}\, e^{i 2\pi
\bolds{\kappa} \cdot
\bolds{\rho}}\,\frac{\tilde\theta(\bolds{\kappa})}{\kappa_x+i\kappa_y},
\eea
where $\bolds{\kappa}=(\kappa_x,\kappa_y)$ is a spatial frequency, $h_0$ is an
arbitrary constant, and we introduce a complex function
$\theta(\bolds{\rho})=\theta_x+i\theta_y$, with Fourier transform
$\tilde\theta(\bolds{\kappa})$. This integration method implicitly assumes
that the spatial support of $\theta(\bolds{\rho})$ is finite. That is, we
assume that the average global tilt of the sample is zero. In practice one can
always balance and zero-pad the recorded quadrant images to satisfy this
condition~\cite{Bon_2012}. Global tilts given by $\overline\theta_x$ and
$\overline\theta_y$, should they exist, may be derived separately from the
average relative intensities of the quadrant images and found by direct
integration of Eq.~(\ref{eq:tilt_gradients}), leading to a baseline surface
tilt $h^{\rm{base}}(\bolds{\rho})-h_0=\overline\theta_x x +\overline\theta_y
y$. This tilt may then be added to the local profile variations found by
Eq.~(\ref{eq:h_basic}), leading to a full solution for $h(\bolds{\rho})$.

In a previous publication~\cite{PAW} we concentrated on high resolution
imaging at the diffraction limit defined by a relatively large
$\rm{NA}_i=0.45$. In this work, we focus instead on another feature of PAW
imaging, namely its capacity to readily image over large FOVs. For this, we
will employ small $\rm{NA}_i$'s. This has the advantage of providing large
depth of fields (DOFs), meaning that surface height variations
$h(\bolds{\rho})$ can be measured over long ranges in a single exposure (i.e.
without having to readjust $h_0$). A disadvantage of large FOVs, however, is
that our lateral resolution is likely to be pixel limited rather than
diffraction limited. We must therefore properly account for pixel-induced
spatial filtering of the recorded images, which in turn leads to a spatial
filtering of $\theta(\bolds{\rho})$ in Eq.~(\ref{eq:tilts}). Such filtering is
written as a convolution $\theta^s(\bolds{\rho})={\rm P}(\bolds{\rho})\otimes
\theta(\bolds{\rho})$, where ${\rm P}(\bolds{\rho})$ is the normalized spatial
filter corresponding to a single pixel, and we assume that the total intensity
$I_{\rm tot}$ varies slowly over the scale of a pixel (recall that for pure
phase samples $I_{\rm tot}$ is uniform). For a square pixel of projected size
$p$ at the sample plane, one obtains ${\rm \tilde P}(\bolds{\kappa})={\rm
sinc}(\kappa_x p)\,{\rm sinc}(\kappa_y p)$, where ${\rm sinc}(x)=\sin (\pi
x)/(\pi x)$, leading to the required modification of the kernel in
Eq.~(\ref{eq:h_basic}): $\kappa_x+i\kappa_y\to (\kappa_x+i\kappa_y){\rm
sinc}(\kappa_xp)\,{\rm sinc}(\kappa_yp)$. A similar modification can be found
in Ref.~\cite{Arnison_2004}, though arising from a different consideration of
image shear rather than pixel size.

\begin{figure}[htbp]
\centerline{\includegraphics[width=0.9\columnwidth]{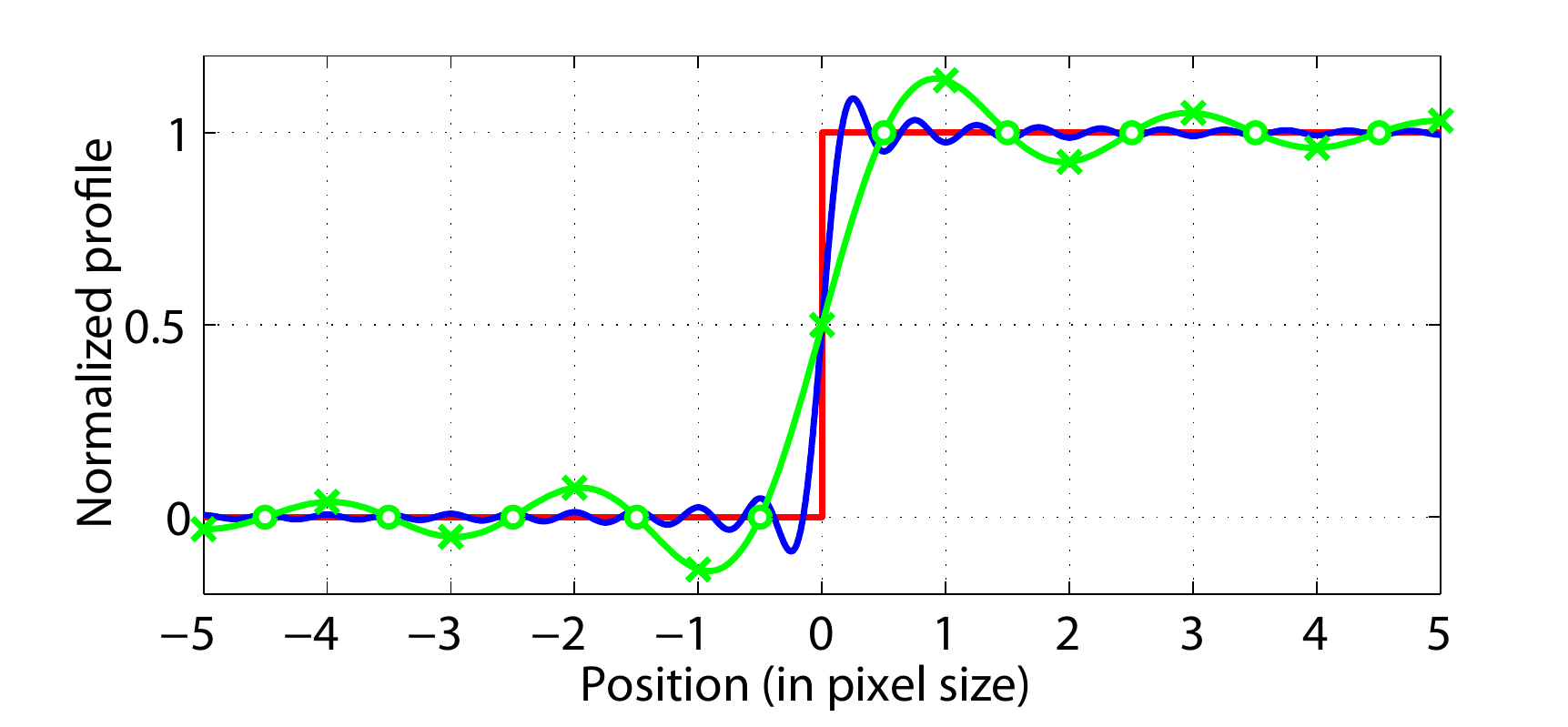}}
\caption{(color online) Step profile discontinuity in sample (red), and
simulated profile reconstruction when system resolution is diffraction limited
(blue) or pixel limited with $\times 2$ undersampling (green). $\circ$ and
$\times$ indicate representative pixel array positions relative to the step
discontinuity.}
\label{fig:step_profile}
\end{figure}

At first glance it might appear that any phase retrieval technique based on
the integration of phase gradients would be susceptible to measurement errors
that propagate upon integration. This is not the case here for two reasons.
First, the spiral phase Fourier integration defined by Eq.~(\ref{eq:h_basic})
dampens the propagation of error (more on this below). Second, sharp phase
gradients that would normally lead to measurement error because they fall
outside our dynamic range (i.e. $|\theta_{x,y}|> {\rm{NA}}_i$), in fact, do
not because they are largely smoothed over by the limited spatial resolution
of our device. As an example, let us consider a sample that features a sharp
phase step as depicted in Fig.~\ref{fig:step_profile}. In the case where the
spatial resolution of our device is diffraction limited ($\approx \lambda /
(2{\rm{NA}}_i)$), the reconstructed phase profile, taking this resolution into
account, exhibits slight ringing but otherwise remains accurate over the scale
of the resolution (provided the phase step is not larger than $\lambda / 2$
over this scale). In the case where the spatial resolution is pixel limited,
the error may or may not be worsened depending on the exact location of the
phase step relative to the pixel array, as illustrated in
Fig.~\ref{fig:step_profile}, but here too the error remains fairly well
localized to within a few pixels of the step. Our phase retrieval method is
thus highly robust.

Our experimental setup is shown in Fig.~\ref{fig:setup}. The illumination
source was an LED with center wavelength $\lambda=525\,{\rm nm}$ and bandwidth
$\sim 20\,{\rm nm}$ (Thorlabs). Two standard Olympus objectives with
magnification 10$\times$ ($\rm{NA}=0.25$) and 1.25$\times$ ($\rm{NA}=0.04$)
were used for medium and low magnification measurements, respectively. The
total system magnification $M_{\rm tot}$ was defined by the microscope tube
lens $f_{\rm{TL}}=150 \,{\rm mm}$, and the PAW module lenses $f_e=150\,{\rm
mm}$ and $f_{\rm{paw}}=50\,{\rm mm}$. $\rm{NA}_i$ was defined by the size of a
square aperture $d_i=3.5\,{\rm mm}$ imaged onto the objective back apertures,
obtaining $M_{\rm tot}=2.8\times$, $\rm{NA}_i=0.097$ and $\rm{DOF}=28.0\,{\rm
\mu} m$ for the 10$\times$ objective, and $M_{\rm tot}=0.35\times$,
$\rm{NA}_i=0.012$ and $\rm{DOF}=1.8\,{\rm mm}$ for the 1.25$\times$ objective.
As a detector we employed a 10-bit machine-vision CCD camera (Hitachi KP-F120)
with square pixels $6.45\,{\rm \mu m}$ in size, capable of acquiring images at
30~fps.

As described, our profilometer cannot distinguish tilt angles $\theta_{x,y}$
induced by the sample from those induced by system aberrations. To correct for
the latter, we first obtained reference quadrant images $I^{\rm{ref}}_k$ from
a simple flat mirror, which, ideally, should be uniform and of equal
intensities. In practice, however, they contained variations due to system
aberrations. This reference measurement need only be performed once for each
setup configuration. Subsequent sample images $I^{\rm{sample}}_k$ were then
normalized according to
\be\label{eq:aberration_correction}
I_k=I^{\rm{sample}}_k/I^{\rm{ref}}_k, \quad k=1..4.
\ee
As noted above, any small misalignment of the reference mirror from normal, as
evidenced by a slight imbalance in the reference quadrature intensities, does
not affect our reconstruction of sample surface variations.

\begin{figure}[htbp]
\centerline{\includegraphics[width=0.95\columnwidth]{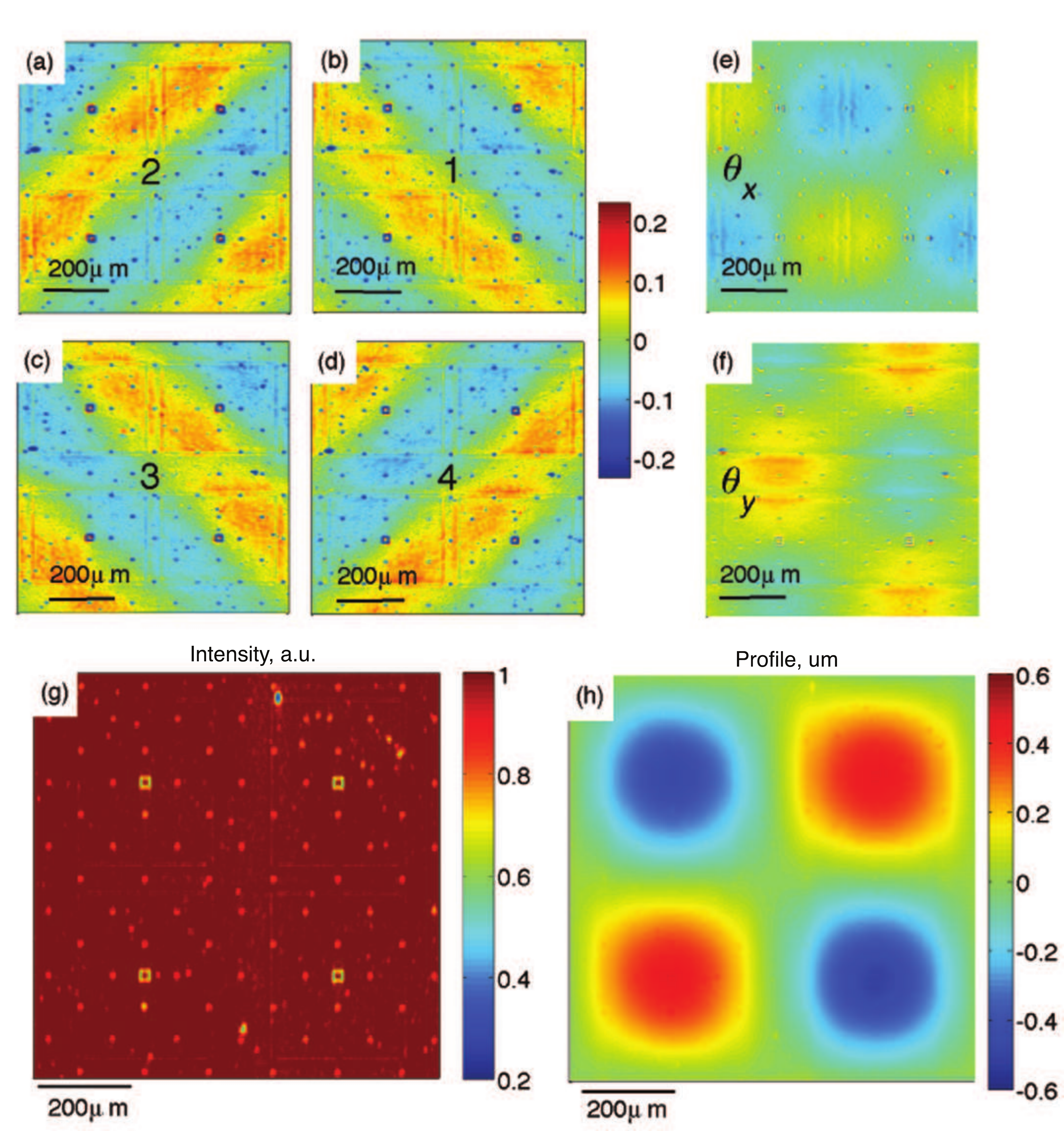}}
\caption{(color online) Images of DM obtained after correcting for system
aberrations according to Eq.~(\ref{eq:aberration_correction}): (a-d) Quadrant
images $I_{1..4}$ (a.u.) recorded by camera, with dc level subtracted for ease
of presentation; (e-f) calculated light tilts $\theta_x$ and $\theta_y$
(a.u.); (g) normalized widefield image $I_{\rm tot}$ (note apparent supporting
structure and etch-access holes); (h) profile of the DM with color-encoded
height ($\mu m$).}
\label{fig:DM_four_images}
\end{figure}

To characterize the performance of our device, we chose a well defined sample
that could be dynamically varied and controlled, namely a deformable mirror
(DM) provided by Boston Micromachines Corp. The mirror consisted of $140$
reflective elements which could be individually controlled at an update rate
of $3\,{\rm kHz}$, forming a grid of period $400\,{\rm \mu m}$ with maximum
stroke about $2\,{\rm \mu m}$. As a test profile, we imposed a two-dimensional
checkerboard pattern. The resulting normalized quadrature images obtained with
the 10$\times$ objective are shown in Fig.~\ref{fig:DM_four_images}.

\begin{figure}[htbp]
\centerline{\includegraphics[width=0.95\columnwidth]{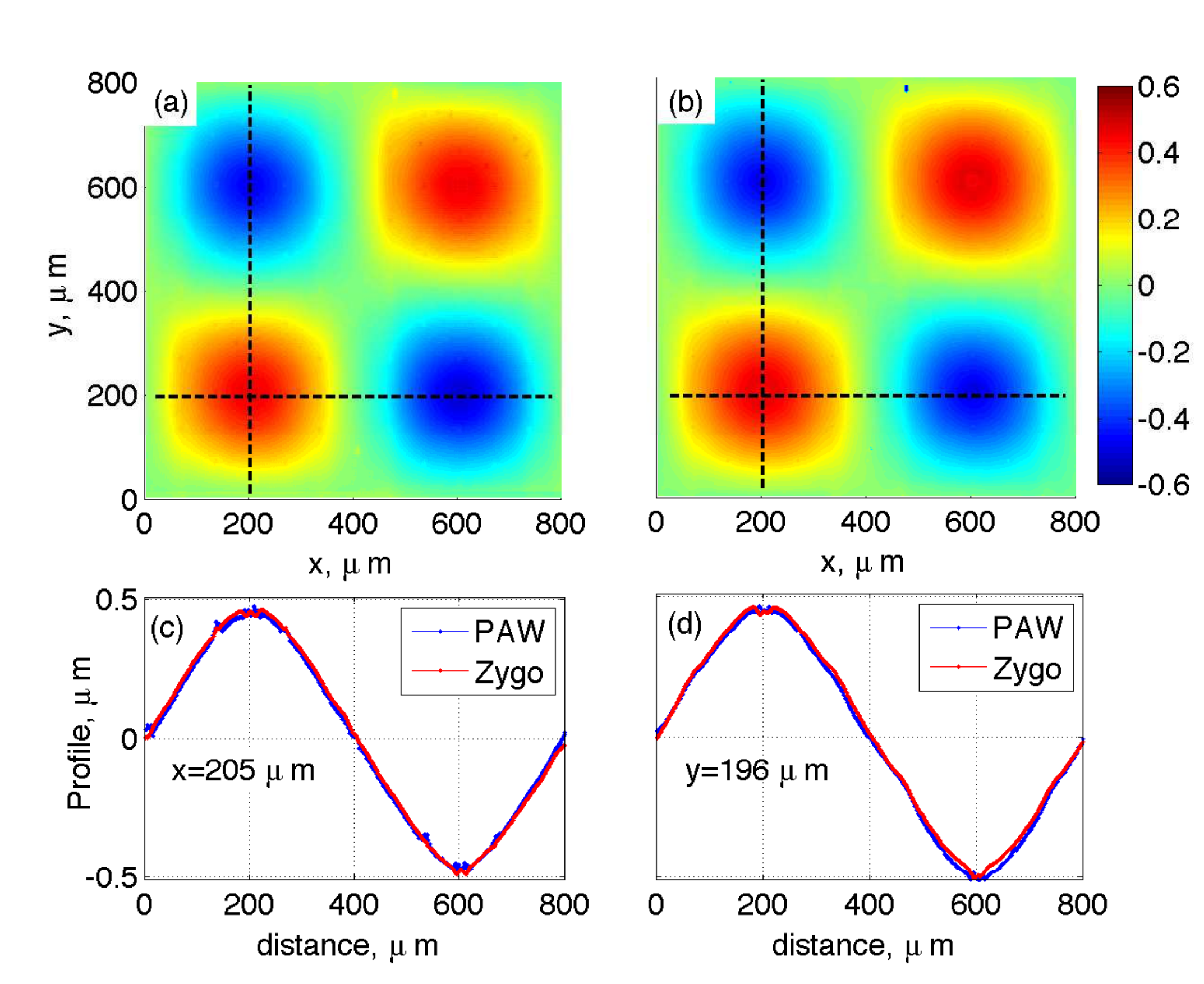}}
\caption{(color online) Profiles of DM reconstructed by (a) PAW and (b) SWLI,
and representative line profiles (c,d) for comparison.}
\label{fig:DM_high_mag}
\end{figure}

To verify the accuracy of our surface height measurements, we compared our
results with those obtained by a commercial profilometer based on SWLI (Zygo
NewView 6300). The results are shown in Fig.~\ref{fig:DM_high_mag}, where the
FOVs from both instruments were scaled and cropped to be identical. The color
bar encodes the reconstructed profile in microns (with average height set to
zero), depicting height variations in the range of about $1\,{\rm \mu m}$. The
agreement between the two techniques is excellent, with root-mean square
discrepancies not exceeding $20\,{\rm nm}$ over the entire FOV. As expected,
we observe the largest discrepancies close to the etch-access holes in the DM
where the surface height variations vary rapidly. Again, these discrepancies
are local only, and can be readily identified from the widefield image $I_{\rm
tot}$.

\begin{figure}[htbp]
\centerline{\includegraphics[width=0.8\columnwidth]{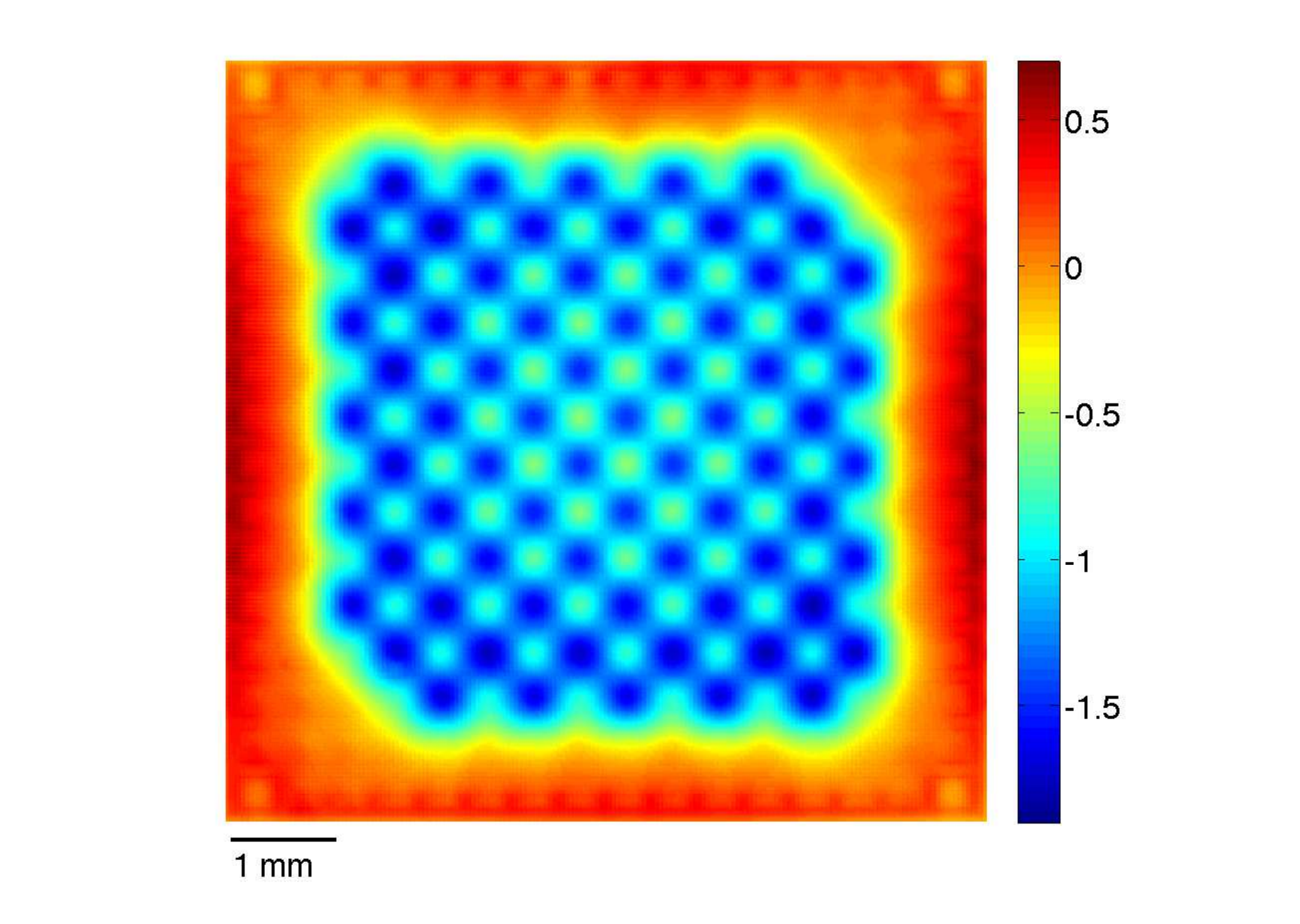}}
\caption{(color online) Profile of entire DM surface (color encoded in $\mu m$) imaged at low magnification and reconstructed by PAW.}
\label{fig:DM_low_mag_2D}
\end{figure}

We note that a key difference between PAW and SWLI profilometry is that PAW
can provide measurements at video rate compared to the several seconds
typically required by SWLI. Another key difference is the ready capacity of
PAW to operate over large FOVs. By simply switching to the 1.25$\times$
objective (and recalibrating), we extended the FOV eight-fold from $800\,{\rm
\mu m}$ to $6.5\,{\rm mm}$. In this manner, we were able to acquire a surface
profile of the entire DM active area, as illustrated in
Fig.~\ref{fig:DM_low_mag_2D}, again at video-rate.

Finally, we consider the noise characteristics of our device. In particular,
we evaluate the uncertainty in our height measurements arising from
unavoidable sources such as shot noise and camera readout noise.
Equation~(\ref{eq:h_basic}) can be recast as $\la
h(\bolds{\rho})\ra-h_0=K(\bolds{\rho})\otimes \la\theta(\bolds{\rho})\ra $,
where $K(\bolds{\rho})=(x-iy)/4\pi \rho^2$ and $\la...\ra$ indicates an
average over multiple measurements. Provided light tilt fluctuations
$\delta\theta=\theta-\la\theta\ra$ are locally uncorrelated, we have $\la
\delta\theta(\bolds{\rho}_1) \delta\theta^*(\bolds{\rho}_2)\ra=
p^2\sigma^2_\theta\delta^2(\bolds{\rho}_1-\bolds{\rho}_2)$, where
$\delta^2(\bolds{\rho})$ is a 2D delta function, and the variance
$\sigma^2_\theta=\la\delta\theta^2(\bolds{\rho})\ra$ is independent of
position $\bolds{\rho}$. This variance can be derived from
Eq.~(\ref{eq:tilts}) and is given by \cite{PAW}
\be\label{eq:sigma_theta}
\sigma^2_\theta=\sigma^2_{\theta_x}+\sigma^2_{\theta_y}=2 {\rm{NA}}^2_i \lb 4\sigma_r^2 +I_{\rm tot}\rb /I^2_{\rm tot} ,
\ee
where $I_{\rm tot}$ is measured in photoelectrons, $\sigma^2_r$ is the camera
readout noise variance, and we assume that the average light tilts are much
smaller than ${\rm{NA}}_i$.

The variance of the surface profile $\sigma^2_h=\la\delta
h^2(\bolds{\rho})\ra$ is thus approximated by
\bea\label{eq:h_variance}
\sigma_h^2 &\approx & \frac{p^2\sigma^2_\theta}{8\pi }\ln (N/2),
\eea
where $N$ is the FOV in pixel counts. The logarithmic dependence on FOV arises
from the long-range behavior of the kernel $K(\bolds{\rho})$. Because this
decays according to a power law, which is largely confined, the dependence on
FOV is weak. From Eq.~(\ref{eq:h_variance}) we expect theoretically
$\sigma_h^{\rm{th}}\approx 3\,{\rm nm}$, which is close to the experimentally
measured $\sigma_h^{\rm{exp}}\approx 2\,{\rm nm}$.

In summary, we have developed an optical surface profiling technique that can
provide video-rate measurements of dynamic samples, featuring
diffraction-limited lateral resolution and nanometric axial resolution
comparable to SWLI. The technique can be implemented as a simple add-on to a
standard lamp-based reflection microscope, making it an attractive and
inexpensive alternative to SWLI.

We thank T. Bifano, J.-C. Baritaux and T. N. Ford for helpful discussions, and
Boston Micromachines Corp. for providing a DM. This work was supported by the
BU Photonics Center through a NSF I/UCRC grant.

\end{document}